\documentclass[a4paper,fleqn,usenatbib]{mnras}
\usepackage{newtxtext,newtxmath}
\usepackage[T1]{fontenc}
\usepackage{ae,aecompl}
\usepackage{graphicx}	
\usepackage{amsmath}	
\usepackage{amssymb}	

\title[Disc Geometry Evolution during Hard States]
{Evolution of Accretion Disc Geometry of GRS~1915+105 during its $\chi$ state as revealed by TCAF solution}

\author[B. G. Dutta et al.]{
Broja G. Dutta,$^{1,2}$\thanks{E-mail: brojadutta@gmail.com}
Partha Sarathi Pal,$^{3}$\thanks{E-mail: parthasp@mail.sysu.edu.cn; parthasarathi.pal@gmail.com}
and Sandip K. Chakrabarti $^{4,2}$\thanks{E-mail: chakraba@bose.res.in}\\
$^{1}$Department of Physics, Rishi Bankim Chandra College, Naihati, W.B, 743165, India.\\
$^{2}$Indian Centre for Space Physics, Kolkata, 700084, India.\\
$^{3}$Sun Yat-sen University, 135 West Xingang Road, Guangzhou, 510275, China.\\
$^{4}$S. N. Bose National Centre For Basic Sciences, Kolkata, 700098, India.
}

\date{Accepted XXX. Received YYY; in original form ZZZ}
\pubyear{2018}
\begin{document}
\label{firstpage}
\pagerange{\pageref{firstpage}--\pageref{lastpage}}
\maketitle

\begin{abstract}
The evolution of the C-type low frequency quasi-periodic oscillations (LFQPOs) 
and associated time lag in transient black hole sources as a function of time 
can be explained by variation of the Compton cloud size in a Two Component 
Advective Flow solution (TCAF).
A similar study of a persistent source, GRS~1915+105, has not been attempted.
We fit the evolution of QPOs with propagatory oscillating shock (POS) solution for 
two sets of so-called $\chi$-state observations and find that the shock steadily recedes with almost
constant velocity 
when QPO frequency is decreasing and the spectrum 
is hardening. The shock moves inward with a constant velocity $v_0=473.0$ cm s$^{-1}$ 
and $v_0=400.0$ cm s$^{-1}$ respectively in these two cases, when the QPO frequency is increasing and the spectrum softens. 
This behavior is similar to what was observed in XTE~J1550-564 during the 1998 outburst. 
The time lag measured at the QPO frequency varies in a similar way as the 
size of the Compton cloud. Most interestingly, in both the cases, the lag switches sign (hard 
lag to soft lag) at a QPO frequency of $\sim 2.3 - 2.5$ Hz irrespective of the energy of photons. 
We find, at very low frequencies $< 1$ Hz, the Comptonizing Efficiency (CE) increases
with QPO frequency and at higher QPO frequencies the trend is opposite. The time lags become mostly
positive at all energies when CE is larger than $\sim 0.85\%$ for both the sources.

\end{abstract}




\section{Introduction}

The Galactic black hole X-ray binary GRS~1915+105 has been persistently bright since 
its discovery in 1992 \citep{rm06}. This source exhibits complex timing and 
spectral variability properties which are different from other compact binary sources and hence 
difficult to interpret within the standard framework. Diverse variability patterns 
\citep{g96, han05, wolt02} in X-rays are classified into 12 separate classes based 
on the nature of the light curves and color-color diagrams.
Each of these variability classes is then reduced into three basic states, 
namely, A, B, and C \citep[see also][]{mrk99, mun99}. According to these authors, 
the state `C' is a low-luminosity and spectrally hard state 
while the states B and A both are high and spectrally soft state. They differ only in luminosity while B state 
is related to high-luminosity and A state is related low-luminosity \citep{wolt02}.
Generally, the state C lasts for a few weeks to months and belongs to the so-called 
$\chi$ variability class, which is equivalent to a standard ``low hard state" (LHS). 
In this state, strong type-C QPOs are observed often evolving with time in the 
range $0.5-10$ Hz \citep{mrk99, tv99, rb00}. 
The complex timing behavior certainly challenges models which predict both the 
X-ray variability and the X-ray spectra \citep{mun01}.
It is suggested that these LFQPOs must be linked to the properties of the accretion disk 
since their centroid frequencies are correlated with the disk flux \citep{mrk99} 
while the emitted luminosity itself varies, for example due to changes in mass accretion 
rate \citep{tp99, c10}. In \citet{cm00, v01b, rao00a, skc05} it was shown that the Comptonized photons 
produce QPOs and thus they are intrinsically related to spectral states. 

There are several models in the literature which attempt to explain origins of QPOs.
A possible origin of the type-C QPOs 
could be Lens-Thirring 
precession of the Comptonizing medium, due to misalignment of the black hole spin and 
the inner accretion flow \citep{sv99, svm99, i09}. \citet{h15, m15} confirmed that 
the QPO amplitude depends on the inclination of the binary orbit, strongly suggesting a 
geometric origin \citep{s06}. 
\citet{iv15} studying QPO phase-resolved spectra, found that the iron 
line equivalent width changes over a QPO cycle in GRS~1915+105, strongly pointing to a geometric origin.
Meanwhile \citet{msc96, skc04, ggc14, skc15} suggested that these QPOs
could be related to the resonance oscillation of the Compton clouds since their occurrences
strongly depend on the accretion rates which in turn change the cooling time scale. This latter 
solution was demonstrated by complete hydrodynamic simulations and does not require a spinning
black hole.

The energy dependence of QPOs in GRS~1915+105 has been studied by \citet{cm00, q10, y13}. 
\citet{cm00} found that the RMS amplitude increases with energy. \citet{y13} also found a smooth 
evolution of the dependence of QPO frequency on photon energy. In XTE~J1550-564, for frequencies 
above $\sim 3.2 $ Hz, \citep{l13} found an increase in QPO frequency with energy and below $\sim 3.2$ Hz 
no variations with energy were observed.
The energy dependence of time/phase lags has been studied by \citep{c99, rb00, q10, mp13}. They found a 
smooth relation between the time/phase lag and energy in GRS~1915+105. A similar behavior is also found 
in XTE~J1550-564 \citep{w99, c00}. \citet{q10} observed that the slope of this energy dependent curve is 
positive when hard lags are observed and the slope is negative when soft lags are observed 
with the lag changing sign from positive to negative at $\sim 2$ Hz. The existence of such a crossover
 frequency is of particular interest. 
In this paper we show that in GRS~1915+105 this frequency remains the same for 
different intervals of $\chi$ variability class observation 
(hard state) of GRS~1915+105. In the case of XTE~J1550-564, the lag switches at a QPO frequency $\sim 3.2$ Hz 
\citep{dc16} which is also unique. This unique feature, which appears to 
characterize high inclination galactic black hole candidates (GBHs)
can be understood only through proper understanding of the flow geometry.
The time lags were first explained to be due to the Comptonization of soft 
seed photons by hot electrons, known as `Compton reverberation' \citep{p80, m88} which 
naturally produces hard time lags. Several models are proposed \citep[see][for review and 
references therein]{c99, p01} to explain the hard and soft lags associated with QPOs 
observed in Galactic binary systems. 
According to propagating perturbation model \cite{bl99, l00}, it is not clear how the propagation 
direction changes when the QPO frequency is close to the crossover frequency 
when the phase lag of the QPO changes sign. 
\citet{n11} reproduced the observed time lags in the continuum during the ``plateau'' 
regions of GRS~1915+105 by invoking a temperature stratification in a corona and assuming 
that the optical depth of the Comptonizing region increases as the disk inner radius 
moves inward. 
In GRS~1915+105,\citet{v13} pointed out that self-obscuration effects for the high inclination \citep{mr94} source 
could play an important role in the observed variation of X-ray flux. 
The reason why these models may fail to explain 
this behavior in totality is that they do not account for the fact that the time/phase lags could be due
to multiple origins \citep{dc16} and what we are observing is the resultant of many such causes.

The inclination of the source could be an important parameter while studying the properties of QPO and time lag. 
Using Monte-Carlo simulations, \citet{g11} found that for the same disk flow parameters, the spectrum is hardened 
when the inclination angle is increased. This was later confirmed by \citet{m13} also. \citet{h15} took 
12 transient LMXBs (excluding GRS~1915+105) and considered hard and hard-intermediate states, and found that 
in higher inclination systems, there is systematically harder X-ray power-law emission. Thus, the 
inclination affects the ratio of the hard and soft photons, time lag and QPO amplitude.

Despite these efforts to explain the LFQPOs and their evolution and origin of the lags, 
none appears to be able to explain the cross over of lags for a particular source at the fixed 
QPO frequency, such as, $\sim 2.3-2.5$ Hz in GRS~1915+105 or $\sim 3.2$ Hz in XTE~J1550-564 both of which have 
a similar (high) inclination angle \citep{o11}. 
Earlier, the Two Component Advective Flow (TCAF) solution was used to explain the QPO behavior 
(\citet{rao00a, skc09, d10, n12, mdc16, mcd17}, etc.), spectral properties 
(\citet{shs02, w02, cr04, p06, sds07, cs13, dmcm15, mdc14, cdc16}, etc.), CE variation 
\citep{p13, p14} and time-lag properties \citep{dc16} in BH binary systems during outbursts. 
In TCAF paradigm which originates from self-consistent transonic flow solution, 
the equatorial plane has a standard Keplerian disk and the sub-Keplerian (i.e., lower angular momentum) 
halo surrounds it. CENtrifugal pressure supported BOundary Layer or CENBOL is also formed between 
inner and outer sonic points which plays the role of the Compton cloud.
Only in TCAF, the Compton cloud (i.e., CENBOL) is self-consistently reduced in size as the soft state is approached 
and QPO frequency is increased, and thus the time lag properties are self-consistently linked to the
spectral properties. Other models, such as the the disk corona model \citep{hm93, z03} and the lamp-post models 
\citep{mm96, mf04} are yet to address all the observed complex aspects within a single framework..

In the POS model \citep{skc08, skc09, d10, ggc14}, the resonance oscillation of the 
post-shock region, i.e., the Compton cloud, causes the observed low-frequency quasi-periodic oscillations (QPOs). 
The evolution of QPO frequency is explained by the systematic variation of mean size of the Compton cloud,
i.e., the steady radial movement of the shock front. It moves inwards when the CENBOL is cooled down. 
Opposite is true when it is heated \citep{mcd15}. As the shock moves in with the increase in disk accretion 
rate, the density and optical depth also generally increases. Thus, analysis of the energy-dependent 
temporal properties in different variability timescales can diagnose the dynamics and geometry of accretion 
flows around black holes. This has been shown for the high inclination transient black hole source 
XTE~J1550-564 during its 1998 outburst and the 
low-inclination black hole transient source GX~339-4 during its 2006-07 outburst using RXTE/PCA data \citep{dc16}. 
However, the dependence of lag on photon energy is more intriguing. GX~339-4 was found to exhibit only
the positive lag for all QPO frequencies, while in XTE~J1550-564, the lag switches sign and becomes negative 
at a crossing frequency of $\sim 3.2$ Hz \citep{dc16}. It was explained qualitatively by adding up 
variations of the lag 
components at different length scales \citep{dc16}. The lag is observed in one band with respect to the other band. 
But in each energy band the photons are coming from different physical processes such as Compton scattering, 
reflection, focusing by curved space time, even due to down-scattering etc. 
Therefore the observed lag will depend on the relative contribution of the different processes in that energy bin.
In a soft-lag, the soft photons are delayed due to the longer path
or the delayed hard photons are down scattered to softer energies. 
In a high inclination system the soft lags might also be due to a dominant reflection component.
This effect is opposite in a low inclination system.
Thus the delay will depend on the inclination angle \citep{ccg17,dc16}. 
The dependence on QPO frequency is a consequence of the frequencies being 
dependent on CENBOL size which in turn determines the fraction of photons which can reach the observer
when inclination is high (obscuration effect).
At a given CENBOL size scale the lag changes its sign. This frequency is not universal as the 
dominating lag would depend on the inclination angle and the mass of the black hole which determines 
the length scales of the Compton cloud. 

In case of the variable source GRS~1915+105 which is continuously active in X-rays since its discovery, 
\citet{p11} observed that the average size of Comptonizing region (i.e., CENBOL) 
changes in a well defined way as it transits from one variability class to another. Using Comptonizing
Efficiency (CE) which is the ratio of number of power-law photons to the number of blackbody 
photons in the spectrum at any given instant, they sequence the different classes according to the 
monotonic changes in  $CE$ or geometry. Thus our goal is to study the evolution of temporal
properties of GRS~1915+105 and compare with those obtained in the transient sources.

In the present paper, we extend the earlier studies of \citet{dc16} applied to transient
sources to the persistent X-ray source GRS~1915+105, to understand the 
behavior of its time/phase-lag properties and evolution of accretion geometry by fitting 
QPO evolution with POS solution. We study the evolution of QPO frequency, time lag and CE. 
We specially choose long duration hard state or state C where the variability
properties (such as QPOs and time lag in sub-second scale) can be easily studied. 
We study these properties in GRS~1915+105 and compare with the other high inclination 
black hole candidate (BHC), XTE~J1550-564 which exhibit frequent outburst phases. 
We also calculate the QPO cross over frequency for different energy bands. We interpret the 
results by using the changes in geometry within this hard state as obtained from the
TCAF solution \citep{ct95}.

\section{Observation and Analysis}
 
For this investigation, we chose hard intermediate state ($\chi$ class) data
which is very similar to hard and hard intermediate state of transient sources in respect of QPO
evolution and spectral variations \citep{wolt02, p13}.
We analyzed two different $\chi$ class observations of GRS~1915+105 by {\it RXTE}  in the period MJD 50270 - 50315 
and MJD 50720 - 50760. We termed them as $\chi$ set-1 and  $\chi$ set-2 respectively.
We limit our analysis to observations when low frequency QPOs (LFQPOs) are observed. 
We used Good Xenon, Event and Single Bit modes which contain high time resolution data for timing
analysis, which was performed using the GHATS software\footnote{http://Astrosat.iucaa.in/$\sim$astrosat
/GHATS\_Package/Home.html}, a customized IDL based timing package, which takes care of Poisson noise subtraction and 
dead-time correction \citep{z95} of the FFT data.
We produced Power Density Spectra (PDS) of light curve segments of 16s length in the energy band of 2-30 keV. 
We averaged them to obtain an average PDS for each set observation. The PDSs were normalized
and converted to fractional squared RMS. The power spectra were then fitted with a combinations of 
Lorentzians \citep[see,][]{nm00} using XSPEC v 12.0. We also produce the PDS for 
different energy bands to study energy dependent fractional amplitude variations.

\begin{figure}
\begin{center}
\includegraphics[width=6.0truecm,angle=-90]{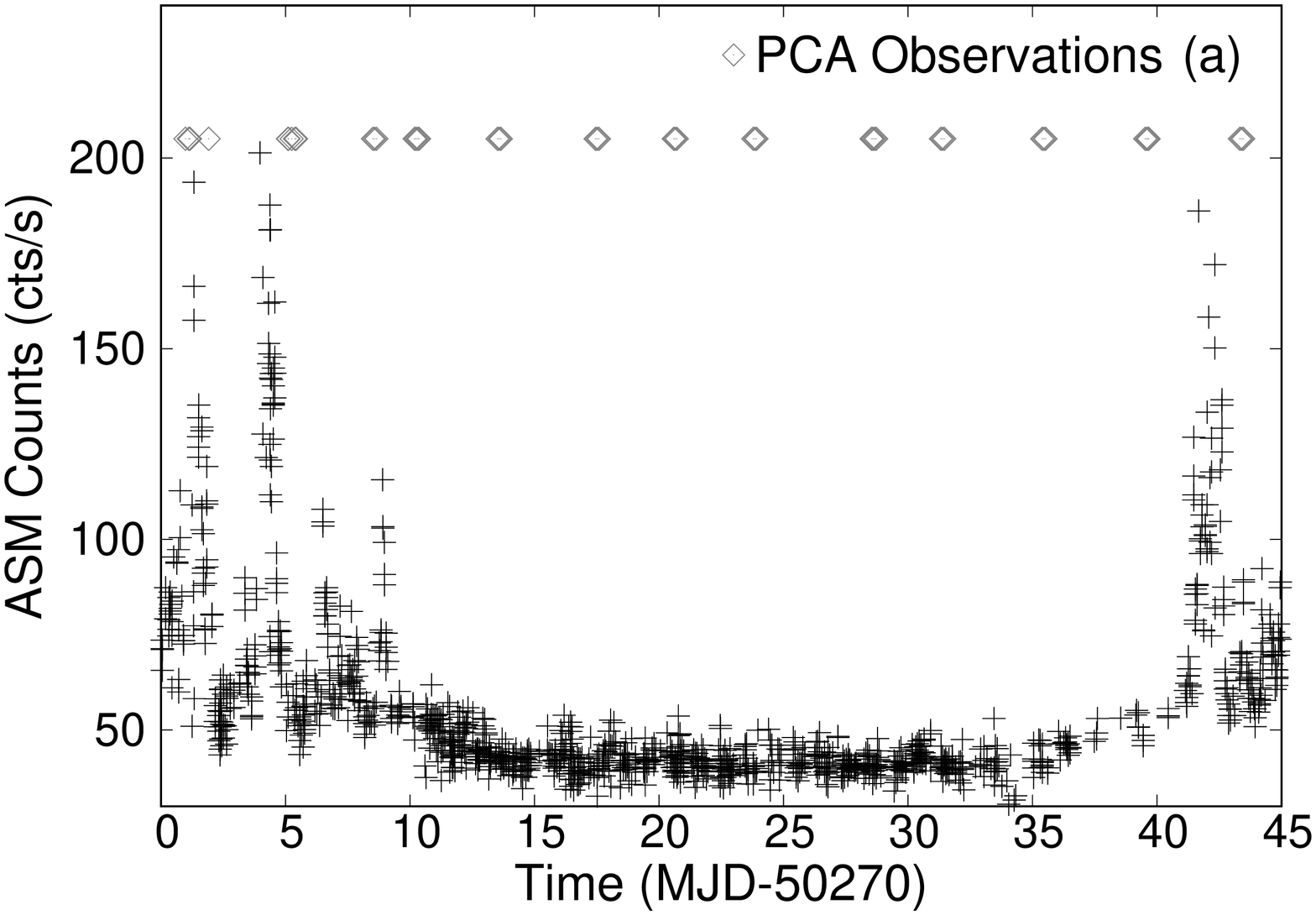}
\includegraphics[width=6.0truecm,angle=-90]{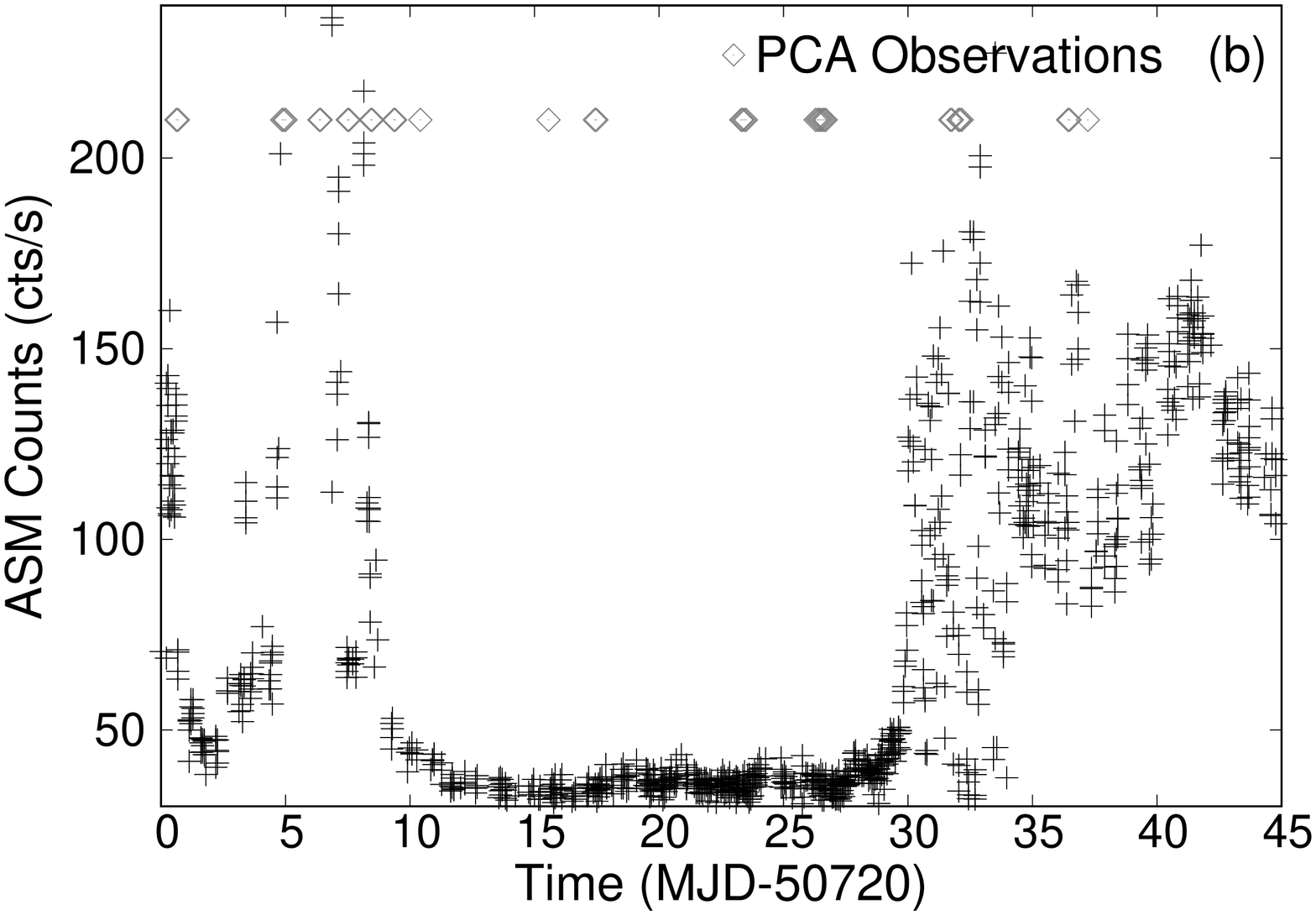}
\caption{(a) ASM light curve during $\chi$ set-1 data, 
analyzed between MJD 50270 and MJD 50315. The gray diamonds 
represent the RXTE-PCA Observations of the days which were available in this period. 
(b) ASM light curve during $\chi$ set-2 data, 
analyzed between MJD 50720 and MJD 50760. The gray diamonds 
represent the RXTE-PCA Observations of the days which were available in this period. 
\label{fig1}}
\end{center}
\end{figure}

Fig.~\ref{fig1}(a) and Fig.~\ref{fig1}(b) shows the ASM light curve for  $\chi$ set-1 and  $\chi$ set-2 
respectively. The gray diamonds in the top of each panel show the dates in which the PCA observation was made.

In our analysis, we produced time lag spectrum for each observation of $\chi$ set-1 \& $\chi$ set-2 data.
We divide the data into a few energy bands depending upon the availability 
of the channel groups. For $\chi$ set-1, the channel groups are: 
0-13 (2.0-5.23) keV, 14-18 (5.23-7.07) keV, 19-25 (7.07-9.57) keV, 26-35 (9.57-13.17) keV, 
36-49 (13.17-18.27)keV, 50-65 (18.27-24.38) keV, 66-89 (24.38-33.06) keV.
For $\chi$ set-2, the channel groups are:
0-13 (2.0-5.23) keV, 14-19 (5.23-7.42) Kev, 20-25 (7.42-9.57) keV, 26-35 (9.57-13.17) keV,
36-50 (13.17-18.64) keV, 51-67 (18.64-24.87) keV, 68-89 (24.87-33.06) keV.
The cross spectrum is
defined as $C(j)=f_{1}^*(j) \times f_{2}(j)$, where $f_1$ and $f_2$ are the complex
Fourier coefficients for the two energy bands at a frequency $\nu_j$ 
and here $f_1^*(j)$ is the complex conjugate of $f_1(j)$ \citep{v87}. 
The phase lag between two band signals at Fourier frequency $\nu_j$ is $\phi_j = arg[C(j)]$
and the corresponding time lag is $\phi_j/{2 \pi \nu_j}$ \citep{u14}. 
We calculate an average cross vector C by averaging complex
values over multiple adjacent 16s cross spectra and then finding the final value of time lag versus frequency. 
Following the method described in \citet{rb00}, 
we calculate time lags at the QPO centroid 
frequency ($\nu_c$) averaging over the interval $nu_c \pm FWHM$ for each energy bands mentioned above 
w.r.t. 2.0-5.23 keV energy band \citep{u14}. In all time lag spectra, positive lag values mean that hard photons are
lagging soft photons. We found that dead time effects are negligible. 

We have already introduced the Comptonizing Efficiency (CE) \citep[see for detail,][]{p14,p13,p11,p15}
which is more accurate than hardness ratios (HR) since it is dynamically computed  
and the energy ranges are dynamically determined from the spectral fit. Normal HR comes from light curve and 
they do not need spectral information. The Comptonizing Efficiency (CE) is defined 
as the ratio of total number of power-law photons ($N_{PL}$) (defined as in
\citet{tit94}) and the total number of soft (seed) photons ($N_{BB}$) from the standard  disc model  of 
\citep{maki86}. The ranges of integration would be different for different spectrum since CE is not just a 
hardness ratio. The limits of energy range we chose is determined case by case as presented 
in \citet{p13} and \citet{p14}.
Thus, the ratio, CE=${N_{PL}}/{N_{BB}}$ can be calculated from the spectrum alone by fitting black body 
and power law component but it has nothing to do with any specific accretion flow model (e.g., TCAF etc.) and 
it does not depend on the mass.

In the TCAF model, the POS solution has been successfully applied to study the evolution of QPO frequencies 
and to explain the variation of accretion 
flow geometry during the rising and the declining phases of the outbursts of a few transient BHCs, such as 
GRO J1655-40 \citep{skc08,skc05}, XTE~J1550-564 \citep{skc09}, GX~339-4 \citep{d10,n12}, H~1743-322 \citep{d13} 
and IGR~J17091-3624 \citep{I15}, MAXI~J1659-152 \citep{mdc16} 
but never applied to any persistent source such as GRS~1915+105. 
In the shock oscillation model of QPOs, the oscillation of X-ray intensity is actually due to the oscillation
of the post-shock (CENBOL) region \citep{msc96, cm00}. The physical reason for the 
oscillation of shocks is as follows: considering a phase during which the shock is receding, in the frame 
comoving with the shock, the upstream flow would appear to be faster, and the post-shock region downstream would
therefore be hotter and would cool down faster. This reduction of post-shock pressure causes the shock to return
towards the black hole,  and the process will continue till it overshoots the equilibrium region and 
comes much closer. From the shock frame the incoming flow would be slower and the post-shock cooling diminishes
and compressional heating pushes the shock outward. Thus these two opposite effects 
cause the shock to oscillate only if the cooling time scale and the compressional heating time scale
(infall time scale) are similar. Thus what we observe is a resonance. Furthermore,
by moving back and forth, the CENBOL expands and shrinks and the number of intercepted photons,
and therefore the processed hard photon number is modulated. With 
a power-law cooling (\citep{msc96}, \citep{skc04}) this was demonstrated.
With a dynamical cooling such as a strong outflow, (e.g., \citet{ry97}) is present, which is true when the
shock condition is not satisfied, the CENBOL oscillates. In outbursting sources, an enhancement of the accretion 
rate increases the local density and thus the cooling rate. The resulting drop of the post-shock pressure 
reduces the shock location and increases the oscillation frequency \citep{mcd15} during the phase 
leading to a softer state. Once the resonance sets in, it would continue to be locked into resonance till 
accretion rates change considerably, causing evolution of QPOs as the object goes to softer and harder 
states respectively. Most interestingly, this condition allows one to obtain QPO frequency from the shock 
location or the size of the Compton cloud (CENBOL).

The location of the shock wave ($X_s$) is calculated from the observed QPO 
frequency ($\nu_{QPO}$) as it is assumed that QPOs are generated due to the oscillations of the shock. 
The generated QPO frequency ($\nu_{QPO}$) is proportional to the 
inverse of infall time ($t_{infall}$, i.e., matter crossing time from the shock location to the 
black-hole), i.e., $\nu_{QPO}  \sim \nu_{s0}/t_{infall}$ where $\nu_{s0} = c/r_s = c^3/{2GM_{BH}}$ is
the inverse of $1~r_s$ light crossing time of a BH of mass $M_{BH}$ \citep{msc96}, \citep{cm00} and also
\begin{equation}
t_{infall} \sim  \, R X_s(X_s-1)^{1/2} \,\sim \, {X_s}^{3/2} 
\label{eq1}
\end{equation}
$R$ is the shock strength ($={\rho_{+}}/{\rho_{-}}$, i.e., ratio of the post-shock to pre-shock densities.)
Here we used the fact that the infall of matter in the post-shock region is slowed down by a factor of $R$.
Thus, the QPO frequency according to this model, is $\nu_{QPO} \sim {X_s}^{-3/2}$. The time dependent
shock location is given by, 
\begin{equation}
{X_s}(t) = r_{s0}\, \pm \,{vt}/{r_g},
\label{eq2} 
\end{equation}
where, $r_{s0}$ is the shock location when $t$ is zero, i.e., on the first day of observation
and $v$ is the velocity of the shock wave. The positive sign in the second term is to be used for an outgoing
shock (when the spectral state is gradually becoming harder) and the negative sign is to be used 
for the in-falling shock (when the spectral state is gradually becoming softer). $X_s$ is in units
of the Schwarzschild radius $r_g = 2GM/c^2$ where, $M$ is the BH mass and $c$ is the velocity of light.

\section{Results}

\begin{figure}
\vskip 1.0cm
\begin{center}
\includegraphics[width=7.0truecm,angle=-90]{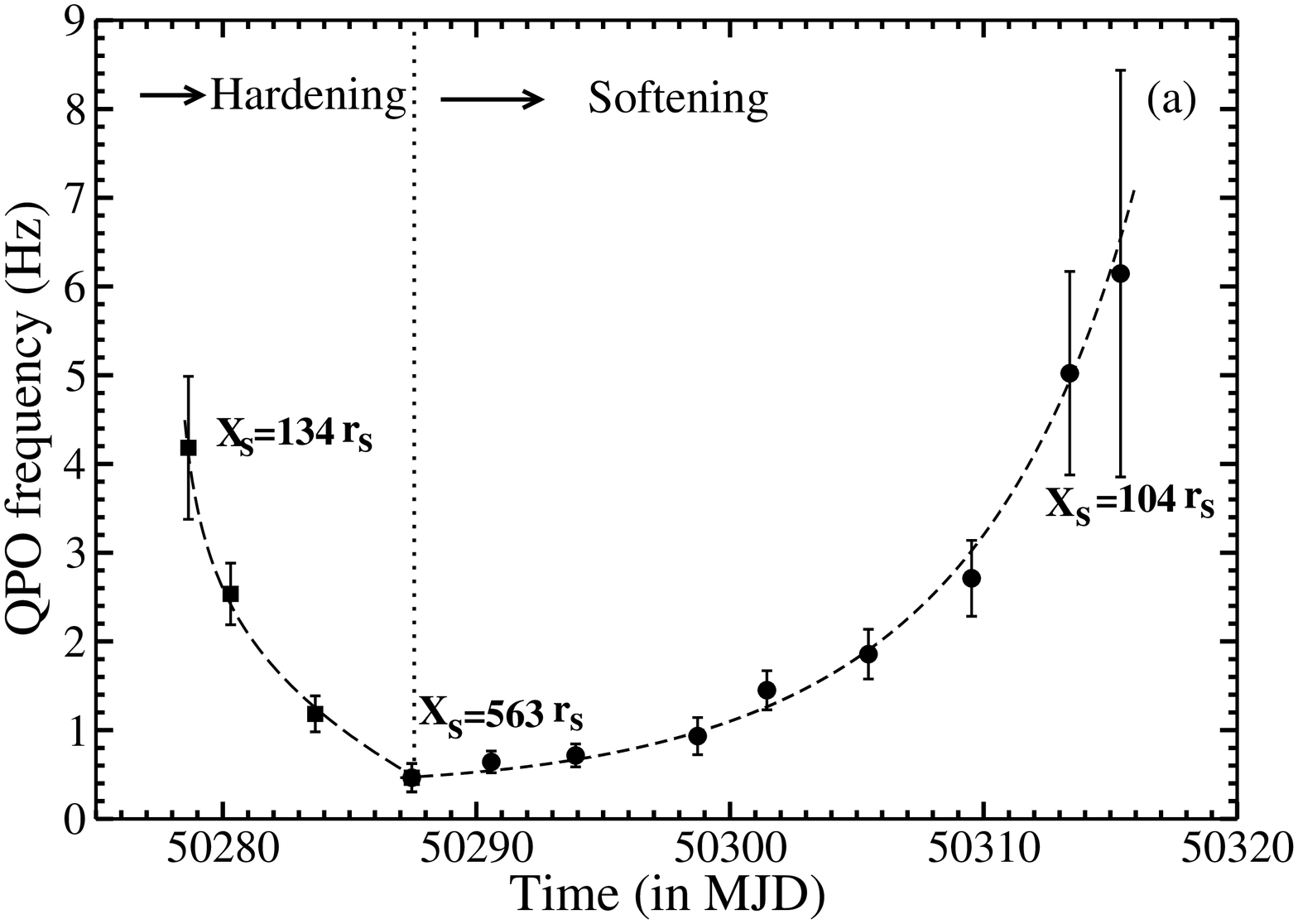}
\includegraphics[width=7.0truecm,angle=-90]{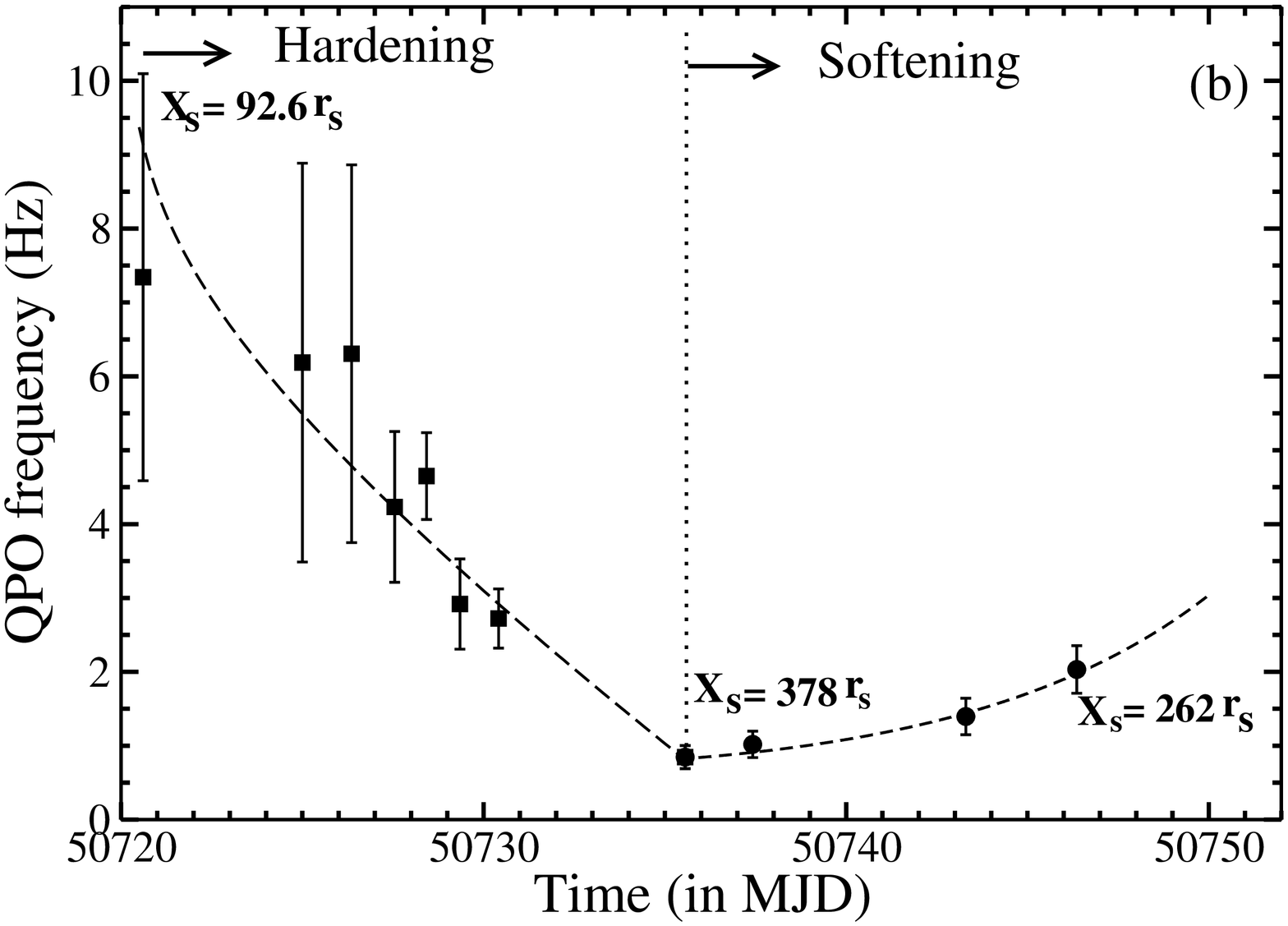}
\caption{ (a) Variations of QPO frequency with time (day) during the hardening and softening phases of $\chi$ set-1
with the fitted POS solution in TCAF model plotted. The shock moves at a constant speed $\sim 473$ cm/s towards the
black hole. (b) Variations of QPO frequency with time (day) during the 
hardening and softening phases of $\chi$ set-2 with the fitted POS solution in TCAF model is plotted.  
Here, the shock moves at a constant speed $\sim 400$ cm/s during softening phase.
The error bars are FWHM of the fitted Lorentzian in PDS. The dotted curve represents POS solution.  
\label{fig2}
}
\end{center}
\end{figure}

The Fig.~\ref{fig2}(a,b) shows the evolution of QPO frequency as a function of days (MJD) 
from MJD 50271 to MJD 50315 and from MJD 50720 to MJD 50757 respectively for the $\chi$ set-1 and $\chi$ set-2. 
As obtained from data, in both the cases we observe the QPO frequency to be decreasing with time when the spectrum is gradually 
hardened and increasing with time after reaching a minimum value when the spectrum is gradually softened. 
The behavior is respectively similar to the declining and rising phases of an outburst.
In $\chi$ set-1 observations, initially (MJD 50271 to MJD 50278), the QPO was stalling around 
$\sim$ $4.18$ Hz and it gradually decreases to $0.46$ Hz (from MJD 50278 to MJD 50287) in nine days. 
Immediately, the QPO frequency starts to increase from $0.46$ Hz 
(MJD 50287) to $6.14$ Hz (MJD 50315.4) in $28$ days. The dashed curve represents our 
fit with POS solutions (equations~\ref{eq1} and \ref{eq2}). We find that the shock is located at  $\sim 134.8 \,r_g$ and it 
increases gradually till $\sim 563.0 \,r_g$ in $\sim 9$ days in the first (hardening) phase and move away 
to $\sim 104.0\,r_g$ in the second (softening) phase.
Depending on the rate of cooling in the post-shock region, which drives the movement of 
shock front, the shock may be accelerating or decelerating. Accordingly, the shock compression ratio $R$ 
also varies as the shock moves in or out in the following way: $1/R \rightarrow$ $1/R_{0} \,\pm \,\alpha (t_{d}^2)$,
where $R_{0}$ is the compression ratio at the first day, $t_{d}$ is the time in days and $\alpha$ is a constant
that determines how the strength of the shock becomes stronger/weaker. $\alpha$ is 
negative when the spectrum is gradually becoming harder and the shock becomes stronger with time 
and positive when the spectrum is gradually getting softer and the shock becomes weaker with time.

In $\chi$ set-1 observations, Fig.~\ref{fig2}(a), the shock was found to drift away with time till MJD 50287 
when $ \nu_{QPO}=0.46$ Hz, $R=1.538$ and the spectrum gets harder. 
It evolves as $ \nu_{QPO} \sim \,t_d^{-0.7}$ and 
$X_s \sim \,t_d^{1.1}$ (since, $ \nu_{QPO} \sim \,X_s^{-2/3}$). After that, the shock steadily moves in towards 
the black hole with almost constant velocity ($v_s \sim \,t_d^{0.1}$) and the spectrum gets softer. 
Similarly, for $\chi$ set-2 observations, Fig.~\ref{fig2}(b), the shock was found to drift away from the black hole
(from MJD 50720) with time till MJD 50735 when $ \nu_{QPO}= 0.845$ Hz and $R=1.538$ (the value is assumed equal to 
that derived from $\chi$ set-1. It evolves as 
$ \nu_{QPO} \sim \,t_d^{-0.9}$ and $X_s \sim \,t_d^{1.4}$. Subsequently, the shock steadily moves in towards 
the black hole with almost constant velocity ($v_s \sim \,t_d^{0.4}$) and reaches at $X_s=390.0 \,r_g$ as the spectrum softens. 
A similar result was also found in XTE~J1550-564 where the shock 
steadily receded away from the black hole with almost constant velocity ($v_s \sim \,t_d^{0.2}$) 
\citep{skc09}. We observed a stalling in the shock motion also in GRO J1655-40 \citep{skc08}.
We find the QPO frequency to immediately increase as the shock front 
starts to drift in the reverse direction i.e., towards the black hole as we observed in XTE~J1550-564 
where we found a sudden increase of QPO frequency towards the black hole \citep{skc09}.

In $\chi$ set-1 observations during the phase when the spectrum gets softer (MJD 50287 to MJD 50315), 
the fit requires the shock to be drifted with a constant velocity $v_0=473.0$ cm s$^{-1}$ towards 
the black hole from $X_s=563.0 \,r_g$ to $X_s=104.0 \,r_g$.
The shock location is observed to be time dependent according to the POS model which explains QPOs and it becomes 
weaker as it propagates towards the black hole. 
$R \rightarrow \,1$ as $r\rightarrow\,r_g$. 
The time needed for the shock to disappear, $t_{d} \sim 28.0$ days and  
using the relation, $1/R\rightarrow 1/R_0 \,+\, \alpha\, t_d^2$, we obtained from fitting that $\alpha = 0.0015$ and $R_0 = 1.538$.
Similarly during MJD 50735 to MJD 50746 of $\chi$ set-2 observations, the fit requires the 
shock (assumed $R_0=1.538$) to be drifted with a constant velocity $v_0=400.0$ cm s$^{-1}$ towards the 
black hole from $X_s=390.0 \,r_g$, with $\alpha = 0.0019$. A similar behavior
was also observed in XTE~J1550-564 during the softening phase of the 1998 outburst where the shock
was propagating with a constant speed $v_0=1981$ cm s$^{-1}$ and $\alpha = 0.022$ \citep{skc09} towards the black hole.

\begin{figure}
\begin{center}
\includegraphics[width=6.0truecm,angle=-90]{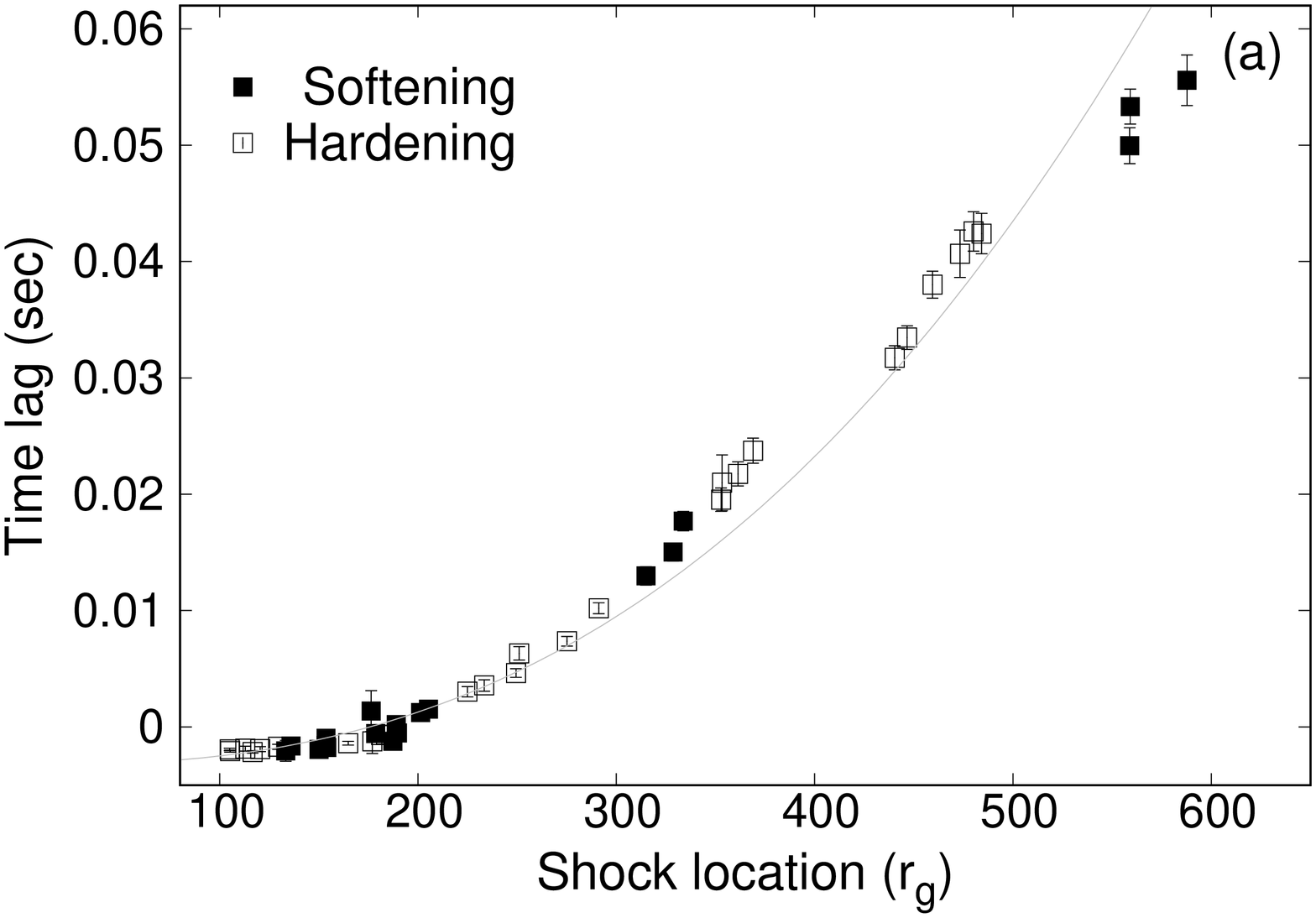}
\includegraphics[width=6.0truecm,angle=-90]{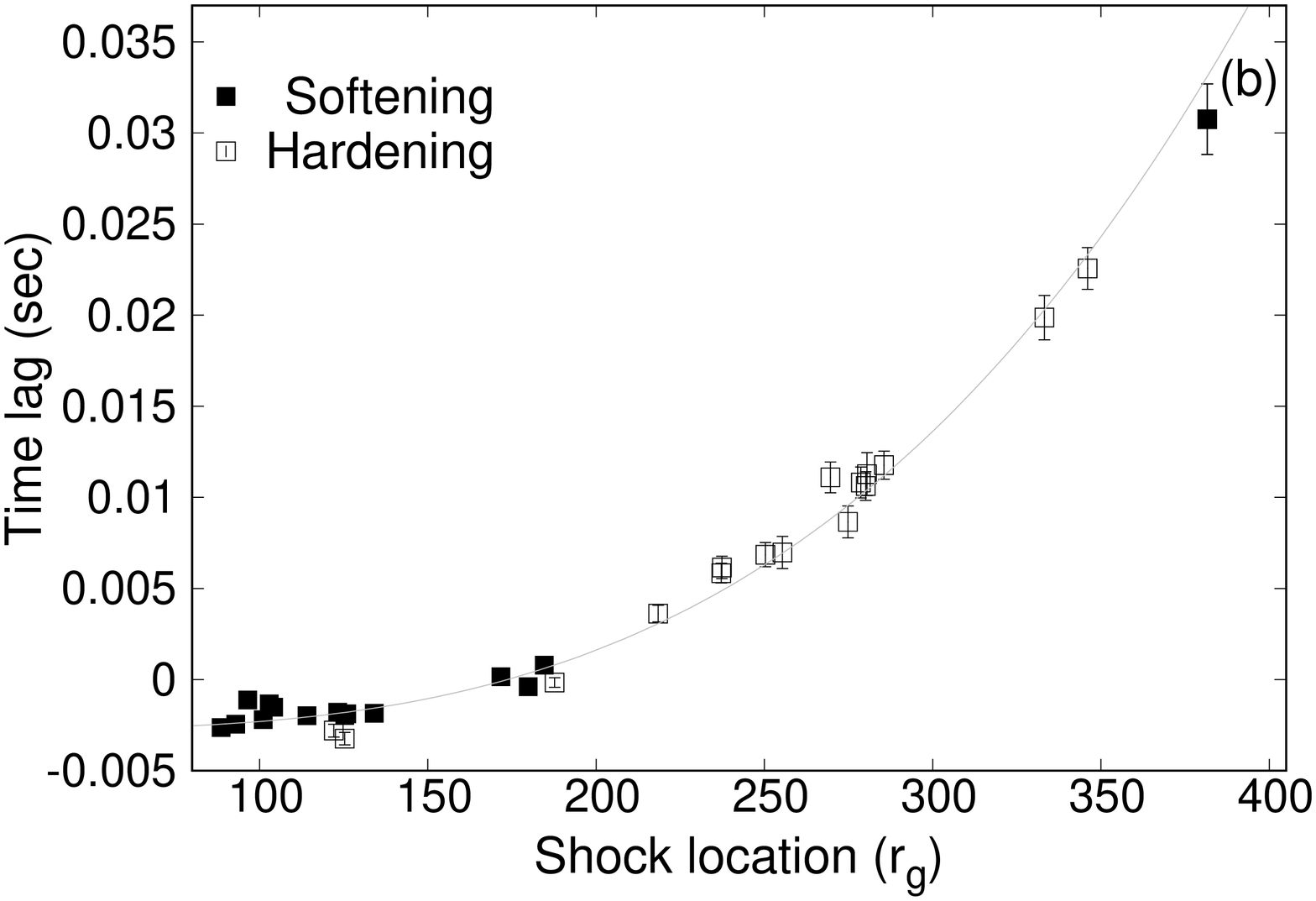}
\caption{Variation of time lag with shock location ($X_s$) is plotted during 
the observations of $\chi$ state. Solid squares represent data during softening and hollow squares 
represent data during hardening. Time lag is calculated at the QPO 
centroid frequency ($f_Q$) integrated over the width equal to the FWHM of the QPO itself. 
(a) $\chi$ set-1 observations and (b) $\chi$ set-2 observations.
\label{fig3}
}
\end{center}
\end{figure}

Fig.~\ref{fig3}(a) shows the variation of time lag of $5.23-7.07$ keV photons w.r.t $2.0-5.23$ keV photons with 
shock locations during the observations $\chi$ set-1. 
Fig.~\ref{fig3}(b) shows the variation of time lag of $5.23-7.42$ keV photons w.r.t $2.0-5.23$ keV photons with 
shock locations during the observations $\chi$ set-2. Solid squares represent the data during the softening phase
of QPO frequency and hollow squares represent the data during the hardening phase of QPO frequency 
for both $\chi$ sets of observations.  
Thus, we find a systematic variation of time lag with shock location (i.e., movement of the shock front)
which in turn represents the variation of Comptonizing region (i.e., CENBOL size) during the rising and 
declining phases of the persistent source for two different $\chi$ sets of observations. 
A similar variation of time lag with shock location was also observed for the transient source XTE~J1550-564 \citep{dc16}.  

\begin{figure}
\begin{center}
\includegraphics[width=6.0truecm,angle=-90]{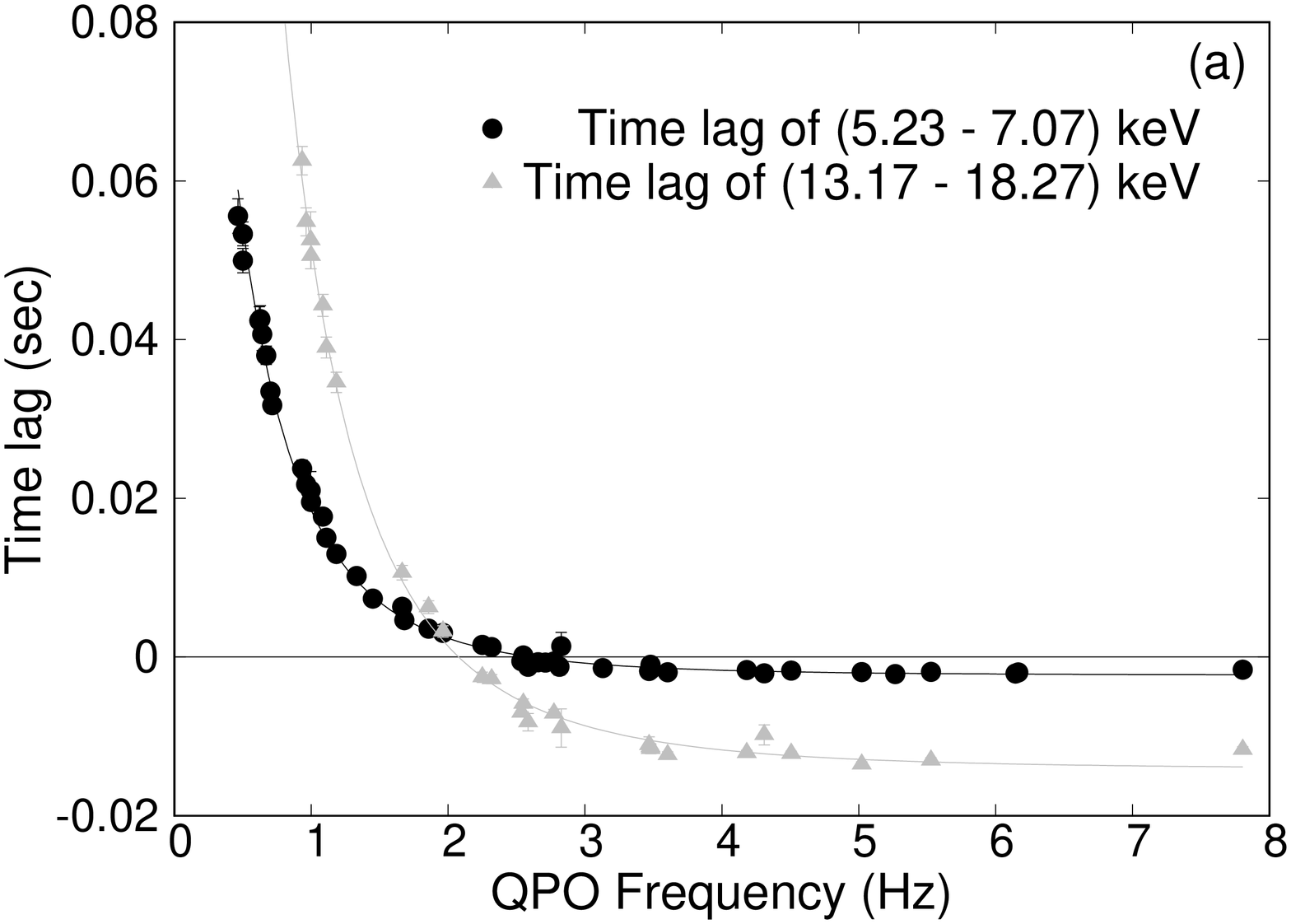}
\includegraphics[width=6.0truecm,angle=-90]{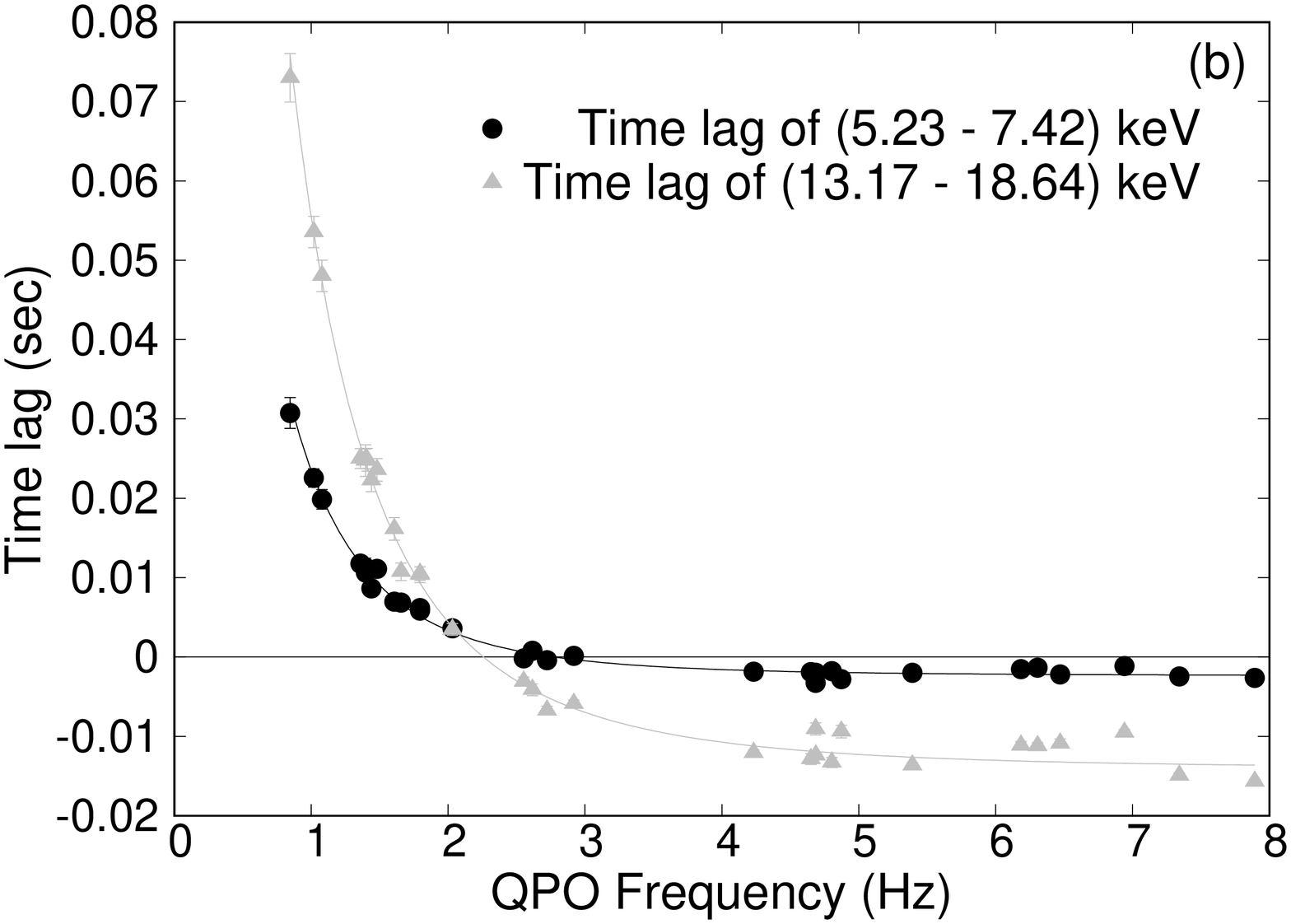}
\caption{Time lag variation with QPO frequencies for different energy bands.
We choose $2.0-5.23$ keV as a reference energy band. 
(a) Black dots represent time lag of $5.23-7.07$ keV 
energy band photons and gray triangles shows that of $13.17 - 18.27$ keV energy band 
photons w.r.t the reference energy band during the $\chi$ set-1 observation. The fitted curves 
show that the time lag cross over frequency at $\sim 2.66$ Hz and $\sim 2.08$ Hz. 
(b) Black dots represent time lag of $5.23-7.42$ keV energy band photons and gray triangles 
represent that of $13.17 - 18.64$ keV energy band photons w.r.t the reference energy band
during the $\chi$ set-2 observation. The fitted curves show that the time lag cross over 
frequency at $\sim 2.77$ Hz and $\sim 2.26$ Hz. 
\label{fig4}
}
\end{center}
\end{figure}

In Fig.~\ref{fig4}(a), we plot the time lags of $5.23-7.07$ keV photons (black dots) 
and $13.17-18.27$ keV photons (gray triangles) w.r.t. $2.0-5.23$ keV photons with QPO frequency in 
the $\chi$ set-1 observations. 
The fit of these curves yield the values of switching frequency $\sim 2.66$ Hz and $\sim 2.08$ Hz 
for the time lags of $5.23-7.07$ keV and $13.17-18.27$ keV photons w.r.t $2.0-5.23$ keV photons respectively.
In Fig.~\ref{fig4}(b), we plot the time lags of $5.23-7.42$ keV photons (black dots) 
and $13.17-18.64$ keV photons (gray triangles) w.r.t. $2.0-5.23$ keV photons with QPO frequency in  
the $\chi$ set-2 observations. 
Again, the fit of these curves yield the values of switching frequency $\sim 2.77$ Hz and $\sim 2.26$ Hz
for the time lags of $5.23-7.42$ keV and $13.17-18.64$ keV photons w.r.t $2.0-5.23$ keV photons respectively.
We will discuss the implication of this in Sect. 4. 

\begin{figure}
\begin{center}
\includegraphics[width=6.0truecm,angle=-90]{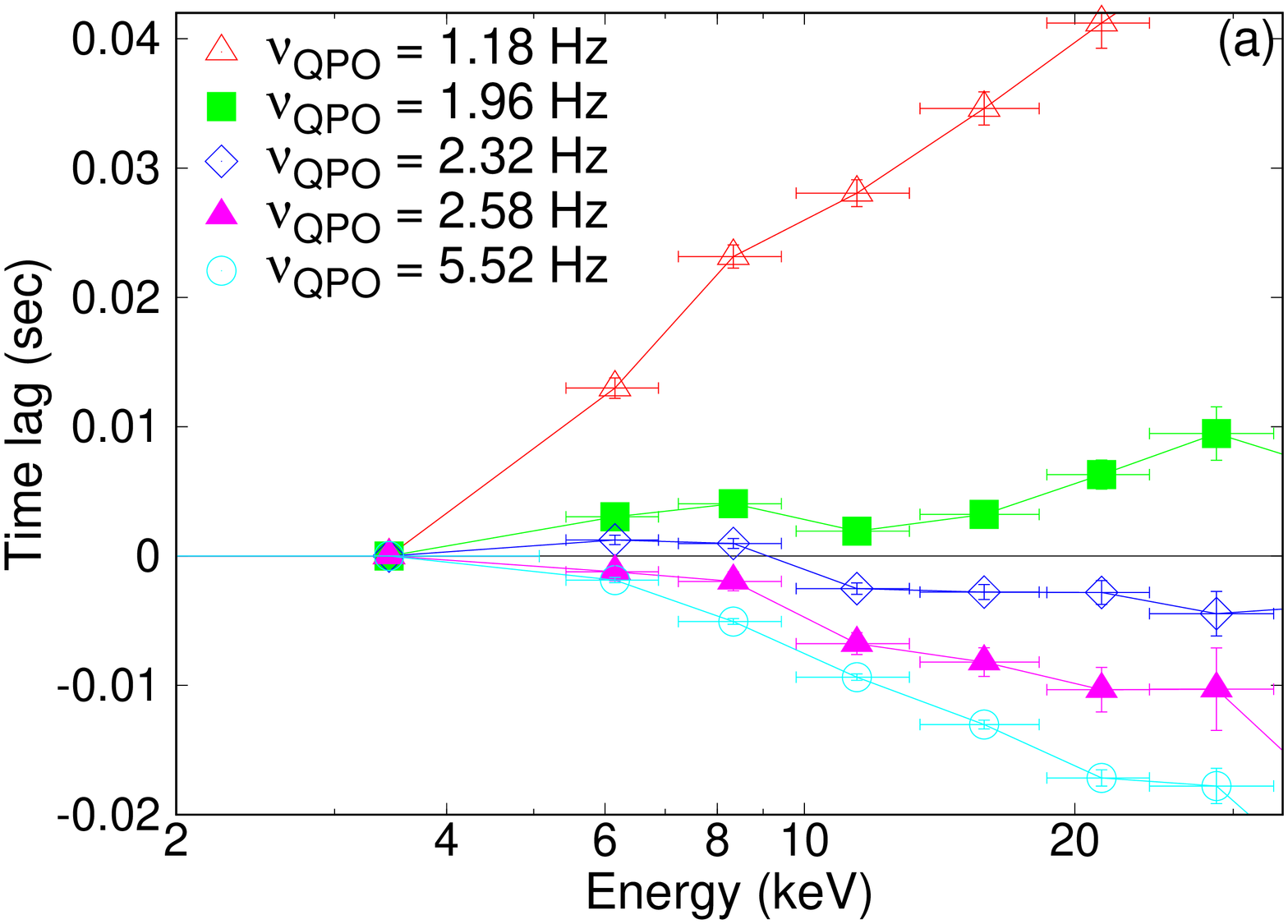}
\includegraphics[width=6.0truecm,angle=-90]{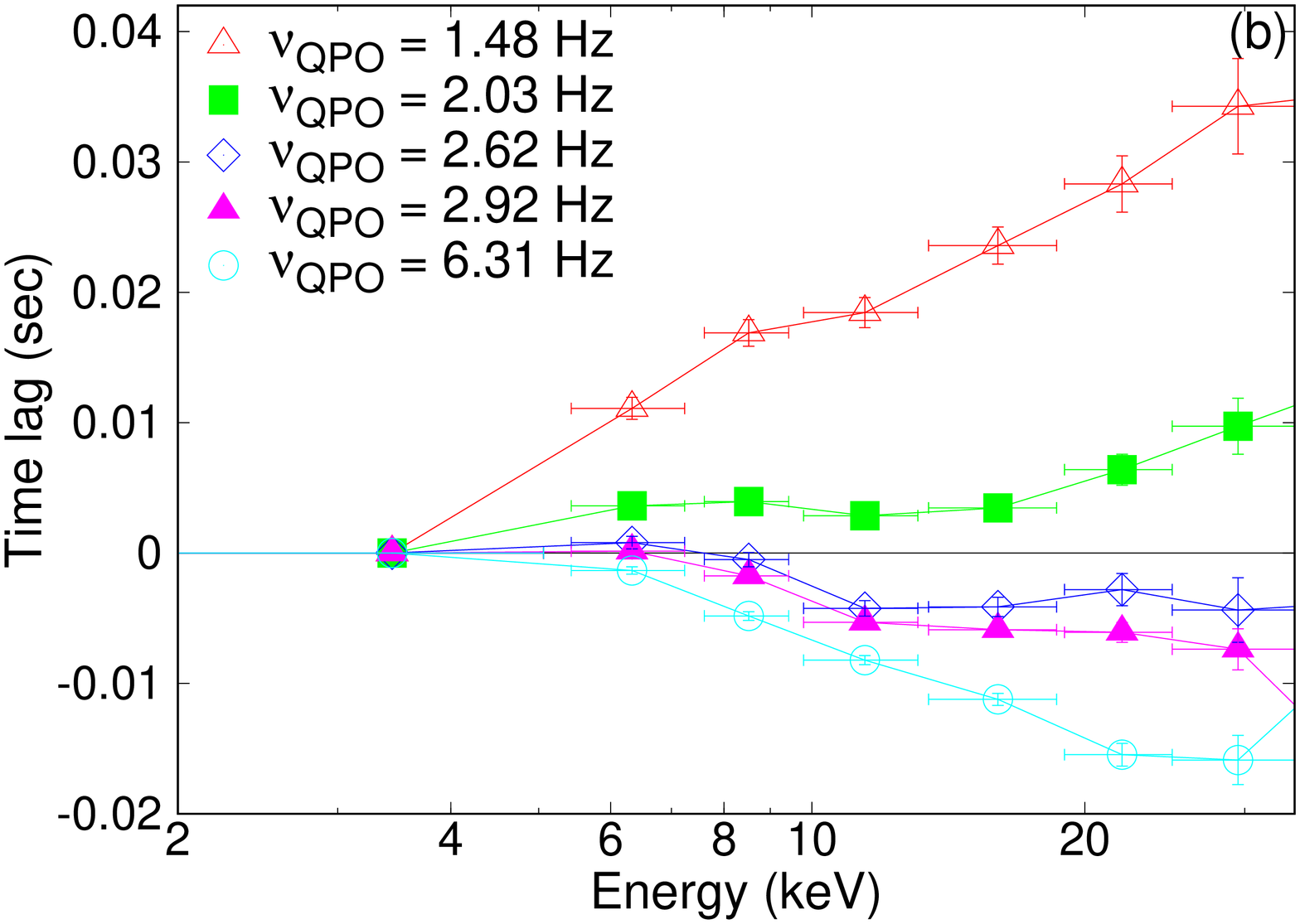}
\caption{Energy-dependent time lag (w.r.t $2.0-5.23$ keV) for different QPO centroid 
frequencies are plotted for $\chi$ set-1 (a) and $\chi$ set-2 (b) observations. 
We averaged the time lag over the width equal to the FWHM of the QPO itself.
We calculated the time lag w.r.t the reference energy band w.r.t. 2.0-5.23 keV.
The time lag switches its sign around QPO frequency of $ \sim 2.32$ Hz (a) and  $\sim 2.62$ Hz (b) for
$\chi$ set-1 and $\chi$ set-2 observations respectively. 
\label{fig5}}
\end{center}
\end{figure}

In Fig.~\ref{fig5}, we plot the energy dependence of time lags for different
energy band photons mentioned in time lag calculation section w.r.t. $2.0-5.23$ keV energy band
at different QPO centroid frequencies ($f_Q$) for both $\chi$ set of observations.
In Fig.~\ref{fig5}(a), we show the energy dependent lag for five different QPO frequencies, viz.
$f_Q=1.18$ Hz (hollow red triangle), $f_Q=1.96$ Hz (solid green square), 
$f_Q=2.32$ Hz (hollow blue diamond), $f_Q=2.58$ Hz (solid magenta triangle) and $f_Q=5.52$ Hz (hollow cyan circle).  
During $\chi$ set-1 of observations for $f_Q \sim 2.32$ Hz, time lag is positive up to $8$ keV. 
We find when $f_Q > 2.32$, time lag becomes negative (i.e., soft lag) at all energies. Time
lag for QPO frequencies below $2.32$ Hz are always positive in all observed energy range. 
 In fig.~\ref{fig5}(b), we plot the same as in fig.~\ref{fig5}(a) for the $\chi$ set-2 observations.
We calculate this energy dependent lag for five QPO frequencies, viz., 
$f_Q=1.48$ Hz (hollow red triangle), $f_Q=2.03$ Hz (solid green square), 
$f_Q=2.62$ Hz (hollow blue diamond), $f_Q=2.92$ Hz (solid magenta triangle) and $f_Q=6.31$ Hz (hollow cyan circle).  
During $\chi$ set-2 observations, for $f_Q=2.62$ Hz, time lag is positive up to $6$ keV. 
Again, we find $f_Q > 2.62$, time lag becomes negative (i.e., soft lag) at all energies. 
The time lags for QPO frequencies below $2.62$ Hz are positive throughout all observed energy
range. Time lag is always negative for the entire energy range when the 
oscillation frequency is greater than $2.62$ Hz.  

In Fig.~\ref{fig5}(a,b), we show that 
the time lag monotonically increases with energy for QPO frequency $\sim 1.18$  Hz, but
monotonically decreases from a frequency of $\sim 2.3$ Hz onward for $\chi$ set-1. Similarly, time lag
monotonically increases with energy for QPO frequency $\sim 1.48$  Hz, but
monotonically decreases from a frequency of $\sim 2.62$ Hz onward for $\chi$ set-2.
A similar type of behavior was also observed for the transient source XTE~J1550-564 \citep{dc16}.
Physical processes producing high-energy photons imprint energy-dependent time lags on the 
output radiation, therefore an energy dependence in the measured lags of high energy photons is naturally expected.
Repeated Comptonization scatterings always produce longer time lags for hard photons. 
Similarly, the gravitational bending of photons (i.e., focusing of emitted photons) is 
energy and geometry dependent since higher-energy photons are expected to come from regions 
closer to a black hole \citep{ccg17,dc16}. Another major effect
is the reflection of higher-energy photons which produce soft lag at high inclination angle \citep{dc16}. 

\begin{figure}
\begin{center}
\includegraphics[width=6.0truecm,angle=-90]{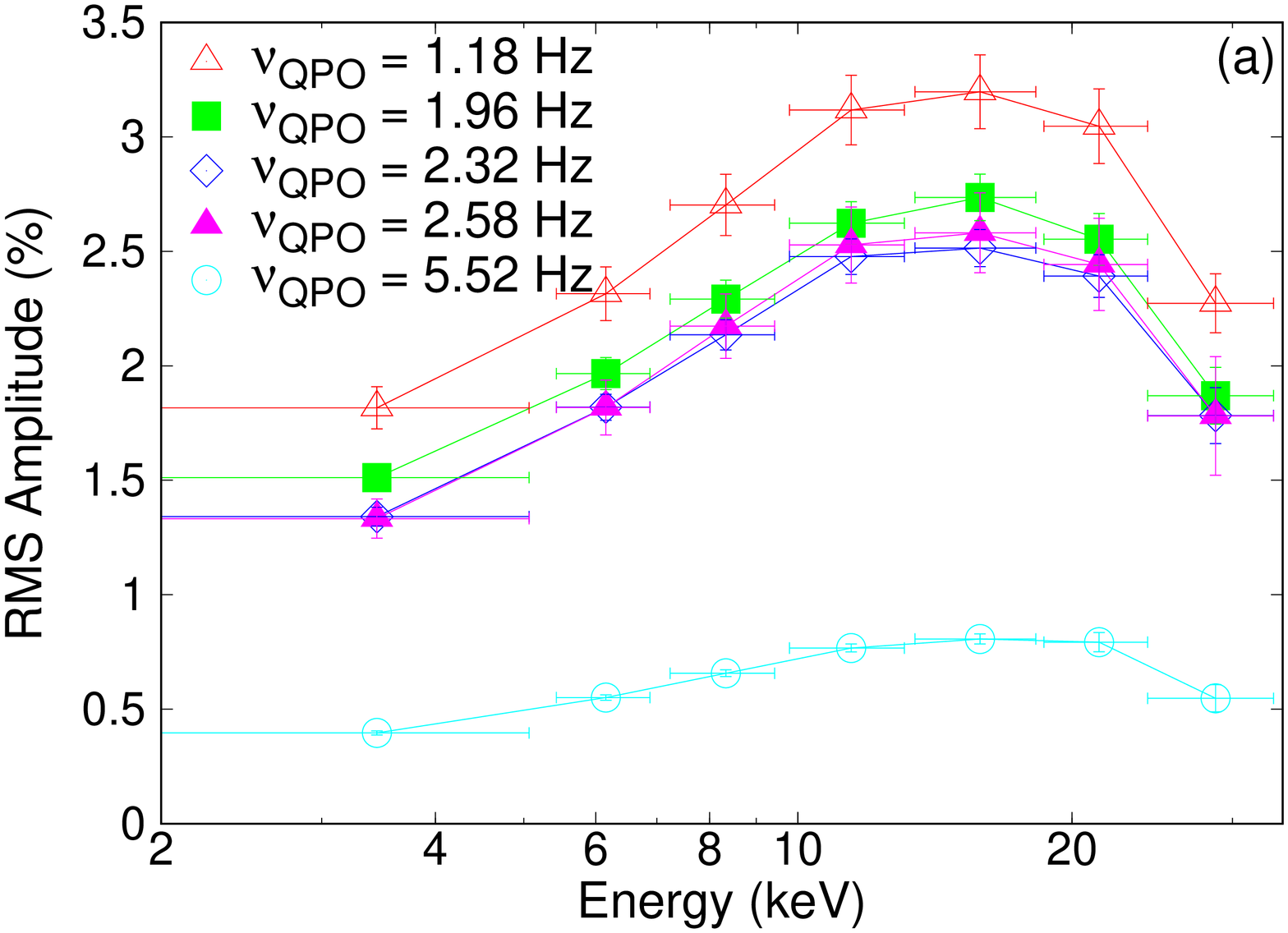}
\includegraphics[width=6.0truecm,angle=-90]{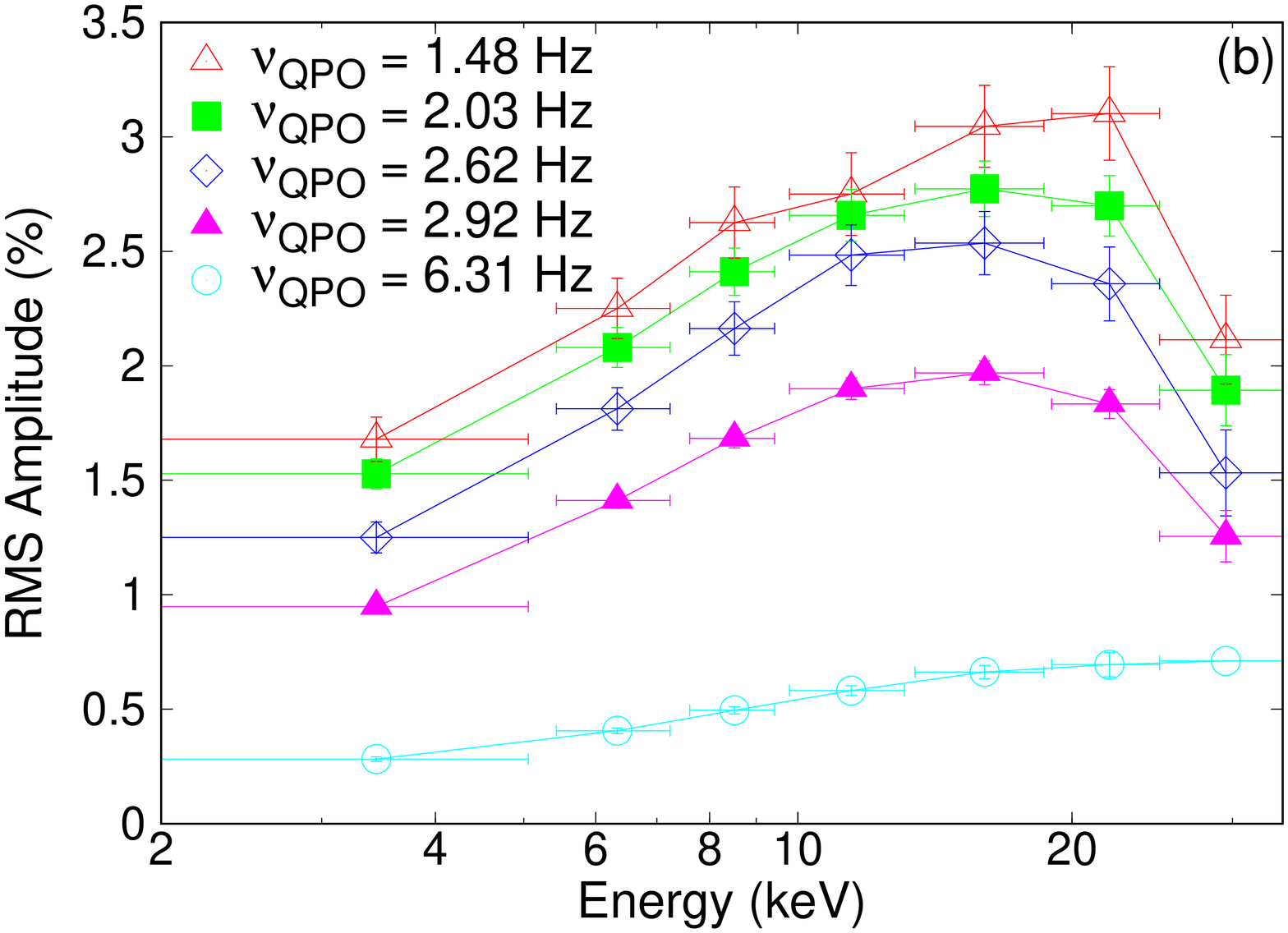}
\caption{(a) Energy dependent RMS amplitude variation at different QPO frequencies. The energy
range used here as in Fig.~\ref{fig5}(a) for $\chi$ set-1 observation.
(b) Energy dependent RMS amplitude variation at different QPO frequencies and the energy range 
used here as used in Fig.~\ref{fig5}(b) for $\chi$ set-2 observation.
\label{fig6}}
\end{center}
\end{figure}

In Fig.~\ref{fig6}(a,b), the variation of fractional RMS amplitude for different QPO frequencies
are plotted for $\chi$ set-1 and $\chi$ set-2 observations respectively. Here, we choose the 
QPO frequencies which are used in Fig.~\ref{fig2} (a,b) and symbols are used as in Fig.~\ref{fig5}(a,b) 
for $\chi$ set-1 and $\chi$ set-2 observations respectively. In Fig.~\ref{fig6}(a), 
we find that the RMS amplitude increases with energy, reaches a maximum and then starts to decrease with energy. 
This implies that for both $\chi$ set observations, the mid-energy photons 
contribute to the quasi-periodic oscillations more than softer photons. 
We have already seen a similar behavior in other Galactic transient sources \citep{rao00a, skc08, skc09,v01b}. 
The rms amplitude of the QPO decreases with increasing QPO frequency because a larger number of photons 
are taking part in oscillation at lower frequencies in comparison to those participating at higher QPO frequencies.
According to the POS model the energy of the peak of the RMS spectrum should decrease with 
increasing QPO frequency, since the spectrum gets softer when the shock is closer to the black hole.
Since the optical depth and temperature of the CENBOL limits
the photon number at the higher end of the spectrum, we expect that the RMS
power would go down at the highest energies. This is what we see in both the data sets.

\begin{figure}
\begin{center}
\includegraphics[width=6.0truecm,angle=-90]{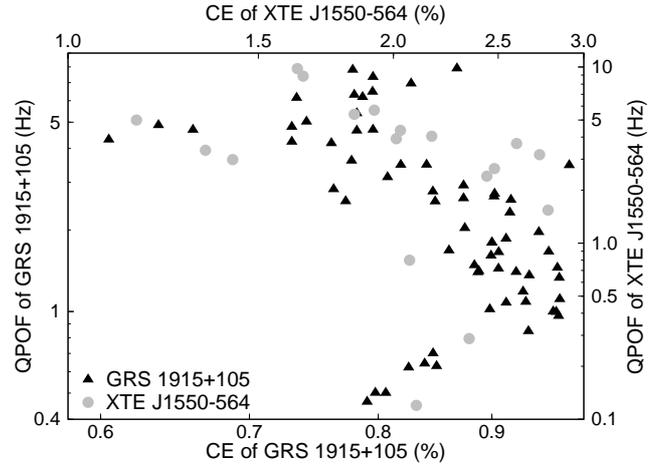}
\caption{Black triangles show the variation of QPO Frequency with CE (\%) 
observed in each of observation during the both $\chi$ set observations. 
Gray circles show the similar variation during 1998 outburst of XTE~J1550-564.
These points are taken from Fig.~6 of \citet{p14}. Their evolutions are roughly along the same track.
\label{fig7}}
\end{center}
\end{figure}

Fig.~\ref{fig7} shows the variation of QPO frequency with the CE. The reason why this plot is important, 
is that the QPO frequency is correlated with the shock 
location and thus, the size of the CENBOL. CE represents the fraction of seed photons intercepted by 
the CENBOL and outflows, if any. During $\chi$ classes when sub-Keplerian halo accretion rate dominates, the 
geometry varies due to the movement of CENBOL boundary (shock
location) and the formation of outflows from it. 
In order to compare the evolution, we superpose the data of the 1998 outburst of 
XTE~J1550-564, reported in Fig.~6 of \citet{p14}. 
We find for both the sources that the CE vs. QPO frequency evolution follows a similar track, 
independently of the fact that one is a persistent source and the other is a X-ray transient.
The two sources have similar inclination (high) angle.
We find for both the objects that the CE vs. QPO frequency evolution follows a similar track 
of evolution and independent of their nature.  

\begin{figure}
\begin{center}
\includegraphics[width=6.0truecm,angle=-90]{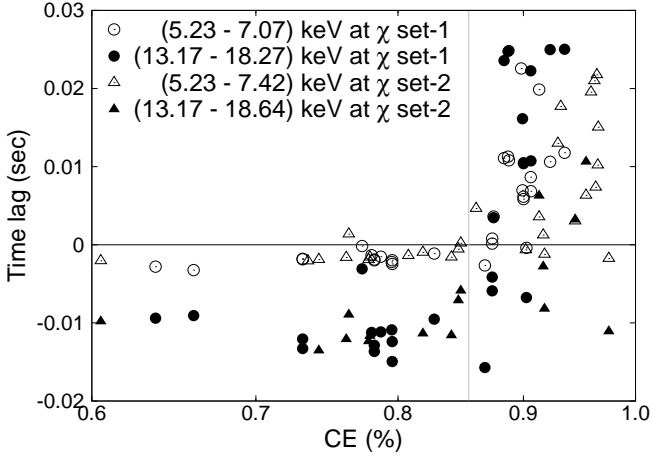}
\caption{Hollow circles and solid circles show the time lag variation with CE (\%)
for (5.23-7.07) keV and (13.17-18.27) keV energy band photons respectively
w.r.t (2.0-5.23) keV energy band photons during $\chi$ set-1 observation.  
Hollow triangle and solid triangle show the time lag variation with CE (\%)
for (5.23-7.42) keV and (13.17-18.64) keV energy band photons respectively
w.r.t (2.0-5.23) keV energy band photons during $\chi$ set-2 observation.  
\label{fig8}} 
\end{center}
\end{figure}

Fig.~\ref{fig8} shows variation of time lag with CE for both $\chi$ set observations. 
In \citet{p11,p13}, it was shown that when the Compton cloud or CENBOL is large
the degree of interception of soft photons with CENBOL will also be greater.
To quantify this, one requires the knowledge of the optical depth of the CENBOL
as well as the fraction of photons intercepted by the CENBOL. 
We see this from our result: at very low QPO frequency $\leq 1$ Hz, i.e., very large CENBOL size, CE is not
at its maximum because the first scattering of photons takes place well inside the CENBOL. Once the CENBOL density
is increased with accretion rates, the CE starts to go down with the CENBOL size.
The larger size of the CENBOL at lower QPO frequencies is also confirmed by the fact that the time lags practically become  
positive at all energies, meaning that Compton up scattering dominates. 
This is seen in Fig.~\ref{fig8} for CE larger than $\sim 0.85$\% and lag $>0$. In the lower
branch $\sim 0.85< $CE $< 1$, the optical depth of CENBOL is not high enough, but the outflow from CENBOL
can also inverse Comptonize the photons, thus producing soft lags.
In the upper branch $CE < 1$, the outflow continues to contribute to the time lag.

\section{Discussions \& Conclusions}

We present a coherent evolutionary picture of the persistent source GRS~1915+105 studying three main 
parameters, the type-C QPO frequency, its associated time lags and the Comptonizing Efficiency, 
in the framework of POS model. We analyzed two different sets of $\chi$ state observations and 
compared our results to those previously obtained from the study of the transient source XTE~J1550-564.
We find the evolution of QPO frequencies as a function of time from MJD 50271 to MJD 50315 and from 
MJD 50720 to MJD 50757 for the $\chi$ set-1 and $\chi$ set-2 data respectively.  
Within POS models, the initial decrease of QPO frequency as a function of time (declining QPO frequency and hardening of
spectrum) is consistent with a shock steadily receding  
with almost constant velocity $v_s \sim t_d^{0.1}$ 
and $v_s \sim t_d^{0.4}$ for the $\chi$ set-1 and $\chi$ set-2 respectively. This
variation is similar to what was observed in XTE~J1550-564, ($v_s \sim t_d^{0.2}$) \citep{skc09}.
Subsequent increase of QPO frequency (rising QPO frequencies and softening of spectrum) 
can be explained in terms of the shock moving towards the black hole with a constant 
velocity of $v_0=473.0$ cm s$^{-1}$ and $v_0=400.0$ cm s$^{-1}$  for the $\chi$ set-1      
and $\chi$ set-2 respectively. In XTE J1550- 564, during a similar spectral variation phase, 
the shock was found to propagate with a much
higher speed $v_0=1981$ cm s$^{-1}$. This difference in velocity could be due to
lower rate of cooling in GRS~1915+105. The fit also requires the shock to be time dependent 
in both the objects, and to become weaker as it propagates towards the black hole. 
Hence, the observed variations of QPO frequency with time resemble the variation in size of the CENBOL.

We find in both the sources that the time lag increases 
when the QPO frequency decreases, i.e., when the size of the Comptonizing region 
(i.e., CENBOL) increases, and that at a specific QPO frequency, i.e., at a 
specific size of the Comptonizing region, the lag changes from hard 
to soft (i.e., switches sign). We also find for both the objects that the CE vs. QPO frequency 
evolution looks similar, independently of the fact that one is a persistent source and 
the other is a transient source. CE represents the fraction of seed photons intercepted by the 
CENBOL and outflows, if any. During the $\chi$ state, when the sub-Keplerian flow rate dominates, the 
geometry varies due to movement of CENBOL boundary and the formation of outflows from it.
At very low QPO frequency $\leq 1$ Hz, i.e., when CENBOL is very large in size, 
CE is not maximum, because the first scattering 
takes place deeper inside the CENBOL where the optical depth $\sim 1$. Then,
the CE decreases with decreasing CENBOL size, i.e., increasing QPO frequency.
The larger size of CENBOL at lower QPO frequencies is also 
confirmed by the fact that the time lags become positive at all energies when CE is larger 
than $\sim 0.85$\% (Fig. ~\ref{fig8}). Again, at very low frequencies $\leq 1$ Hz, we note that the 
CE increases with frequency (Fig. ~\ref{fig7}). This is possible only 
if the optical depth is increasing with decreasing CENBOL size (i.e., when accretion 
rate is also increasing) and/or outflow rate also increases. At higher frequencies, the trend 
is opposite: with the reduction of CENBOL size, the intercepted soft photon flux decreases, as expected normally.
At higher frequencies, CENBOL becomes smaller as the spectrum becomes softer and
the lag becomes negative, particularly for higher inclination sources, because of 
larger effects of reflection and focusing \citep{dc16}.     

The spectrum of the lags at different QPO frequencies shows that for $\chi$ set-1 the hard lags increase monotonically 
with energy up to QPO frequency of $\sim 2.3$ Hz and for $\chi$ set-2, up to a QPO frequency of $\sim 2.62$ Hz.
This is similar to what we found in the transient source XTE~J1550-564 \citep{dc16}.  
This behavior suggests that net observed time lag of the high-energy photons depends on 
the relative contribution of the different physical processes responsible for the high energy emission.
Repeated scatterings and Comptonization always produce longer time lags for higher energy photons. 
The gravitational bending of photons (i.e., focusing of emitted photons) is an energy dependent 
process since higher-energy photons are expected to come from regions closer to a black hole \citep{ccg17,dc16}. 
For instance, high energy photons coming from the inner CENBOL in the opposite side would be focused towards the 
observer, but lower energy photons from farther out will not be focused.
Other major effect is reflection of higher-energy photon which produce soft lags when the inclination angle 
is high \citep{dc16} since the already delayed hard photons from CENBOL would be down scattered at the outer 
part of the disk. We therefore find that depending on the QPO frequency i.e., location of the shock, the lag 
could be positive or negative. 
This behavior suggests that the resultant or net time lag depends on the relative contribution from the different 
physical mechanisms by which they are generated. The fractional RMS amplitude is higher at
medium energies for both $\chi$ set of observations but at very high energies rms goes down
as the number of photons at high energy itself is reduced (at the turn-over energy) 
and also because they come from deep inside the CENBOL where oscillation is not as high and thus do not 
contribute to rms value.

With regard to the evolution of LFQPO properties and CE with time within 
the POS solution, the persistent source GRS~1915+105 and the transient source XTE~J1550-564 do 
not show significant differences.  Just as in the declining phase of an outburst the spectrum is 
hardened and QPO frequency is decreased, the first part of a $\chi$ state also shows exactly 
the same behavior. Similarly, just as in the rising 
phase of an outburst, the spectrum is softened and QPO frequency is increased, the second part of a $\chi$ state
shows the same behavior. The shock location also decreases monotonically in the softening part of the $\chi$ state
and exactly opposite behavior is seen in the hardening part of the state. 
We see that the geometry of the CENBOL evolves in the same way in the two types of sources. More 
generally, this conclusion might apply to any source belonging to the two different classes, with most of the observed 
differences being ascribable to different inclination angles. In other words, the apparently persistent 
class variable source such as GRS~1915+105 behaves as though it is made up of a large number of weak outbursts, 
nature of which is dictated by viscous processes which, in turn, decide the accretion rates.

\section*{Acknowledgments}
We acknowledge the Referee for helpful suggestions to improve the manuscript.
We thank T. Belloni for providing the timing analysis software GHATS.
B.~G.~Dutta acknowledges IUCAA for the Visiting Associateship Programme.
P.~S.~Pal acknowledges the Post-Doctoral Fellowship at Sun Yat-sen University, China.




\bibliographystyle{mnras}
\bibliography{reference} 

\bsp	
\label{lastpage}
\end{document}